\documentclass{emulateapj}

\newcommand{\axp}{4U~0142$+$61}
\newcommand{\spz}{\textit{Spitzer}}

\begin{document}
\bibliographystyle{apj_noskip}

\title{Search for Mid-IR Flux Variations from The Anomalous X-ray Pulsar 
4U 0142+61} 

\author{Zhongxiang Wang and Victoria M. Kaspi}

\affil{Department of Physics,
Ernest Rutherford Physics Building,
McGill University, 3600 University Street, Montreal, QC H3A 2T8, Canada}
\email{wangzx, vkaspi@physics.mcgill.ca}

\begin{abstract}
We report on our \spz\  observations of the anomalous X-ray pulsar 
\axp, made following a large X-ray burst that occurred on 2007 February 7.
To search for mid-infrared flux variations, four imaging observations 
were carried out 
at 4.5 and 8.0 $\mu$m with the Infrared Array Camera from February 14 to 21.
No significant flux variations were detected, and 
the average fluxes were 32.1$\pm$2.0 $\mu$Jy at 4.5 $\mu$m 
and 59.8$\pm8.5$ $\mu$Jy at 8.0 $\mu$m, 
consistent with those obtained in 2005.  The non-detection of 
variability is interesting in light of reported rapid variability
from this source in the near-infrared, 
but consistent with the fact that the source already 
went back to its quiescent state before our observations began, 
as indicated by contemporaneous X-ray  observations. 
In order to understand the origin of the near-infrared variability, 
frequent, simultaneous multi-wavelength observations are needed.

\end{abstract}


\section{INTRODUCTION}

The Anomalous X-ray Pulsars (AXPs) are a small group of isolated, young 
neutron stars having X-ray luminosities greater
than their rotational energy loss rates. They are believed to
be magnetars, neutron stars possessing $\sim$10$^{14}$ G   
magnetic fields, with X-ray emission powered by the decay
of their superstrong magnetic fields \citep{td96}. The magnetar nature 
of AXPs has been strongly supported by their high spin-down rates
and similar radiative properties to soft gamma repeaters, including
short X-ray bursts (\citealt{wt06}; \citealt{kas07}). 
The bursting activity likely reflects
structural changes in the surface magnetic field of
a magnetar.

As the brightest and nearest among the currently known AXPs, \axp\  
has been well studied over wavelength ranges from the X-ray
to mid-infrared (MIR).  Its X-ray emission in the 0.5--10 keV range
can be well described by a blackbody ($kT$= 0.5 keV) plus power law 
(photon index $\Gamma$ = 3.4; e.g., \citealt{pat03}), which is believed to 
arise from the surface and magnetosphere of the star 
(e.g., \citealt*{tlk02}).  Its optical emission 
is pulsed at the 8.7-s spin period (\citealt{km02, dhi+05}) and appears to
have a power-law--like spectrum \citep*{hvk00}, likely also originating from 
the magnetosphere (although no apparent connection can be established between 
the optical and X-ray spectra). A surprise came from the detection of 
the source in the MIR
(4.5 and 8.0 $\mu$m), which revealed a rising spectrum from the near-IR 
(NIR) to MIR \citep*{wck06}. The detection can be interpreted as  
a surrounding debris disk \citep{wck06}. 
The putative disk, presumably formed from fallback material after the supernova 
explosion, is irradiated by the X-rays from the central pulsar and
emits mainly in the MIR. The existence and appearance of such a disk
has been predicted (e.g., \citealt*{lwb91, phn00}).

Recently, optical/NIR flux variability from \axp\ has been 
reported \citep{dv06}. In a total of only 9 observations, significant
flux variations were detected, suggesting such variability is 
very common. This variability is intriguing, since no such flux changes
have been detected in X-rays, even though X-ray observations of the source 
have been made much more frequently than the optical/NIR observations.
An obvious question is whether there is similar 
variability in the MIR. Rapid variability in the MIR would be very
hard to understand in the disk model in the absence of X-ray variability.

On 2007 February 7, a large fast-rise X-ray burst was detected during the
{\em Rossi X-Ray Timing Explorer} ({\em RXTE}) monitoring observations 
of \axp\ \citep{gav+07}. The peak flux of the burst was more than two orders
of magnitude larger than that in the source's quiescent state.
This event provided an opportunity to test the fallback disk model, 
since a MIR flux increase is expected from an X-ray irradiated disk when 
the input X-ray flux is increased (see Figure \ref{fig:disk}). 
For the purposes of searching for variability and testing the fallback
disk model, we observed \axp\  with the {\em Spitzer Space Telescope}.  
In this paper, we report on the results of 
the observations. 

\section{OBSERVATIONS AND DATA ANALYSIS}    
\label{sec:obs}

\subsection{\spz/IRAC 4.5/8.0 $\mu$m Imaging}

We observed \axp\  four times in 2007 February 14--21 with \spz, following
the February 7 burst.
To catch flux variations from the source and possibly
constrain their time scale, the observations were scheduled to be on
days 2, 4, and 7 after the first observation.
The exact observation dates are given in Table~\ref{tab:obs}.
The imaging instrument used was the Infrared Array 
Camera (IRAC; \citealt{fha+04}). It operates in four channels 
at  3.6, 4.5, 5.8, and 8.0 $\mu$m, while two adjacent fields are 
simultaneously imaged in pairs (3.6 and 5.8 $\mu$m; 4.5 and 8.0 $\mu$m). 
We observed our target in the 4.5 (bandwidth 1.0 $\mu$m) and 
8.0 $\mu$m (bandwidth 2.9 $\mu$m) channels.
The detectors at the short and long wavelength bands are InSb
and Si:As devices, respectively, with 256$\times$256 pixels and a plate
scale of 1\farcs2. The field of view (FOV) is 5\farcm2$\times$5\farcm2.
The frame time was 100 s, with 96.8 and 93.6 s effective exposure time
per frame for the 4.5 and 8.0 $\mu$m data, respectively. The total exposure
times in each observation were 19.4 min at 4.5 $\mu$m and 18.7 min at
8.0 $\mu$m.

We also obtained the previous \spz\  IRAC data of \axp\  from the \spz\ archive.
The observation was made on 2005 January 17 at the same wavelength bands as 
ours.  The effective exposure times were 77.4 and 74.9 min in 
the 4.5 and 8.0 $\mu$m, respectively.  The results from this observation 
were previously reported by \citet{wck06}.

\subsection{Data Analysis}

The raw image data were processed through the IRAC data pipelines 
(version S15.3.0) at the {\em Spitzer} Science Center (SSC). 
In the Basic Calibrated Data (BCD) pipeline, standard imaging data
reductions, such as removal of the electronic bias, dark sky subtraction,
flat-fielding, and linearization, are performed and individual flux-calibrated 
BCD frames are produced.  In the post-BCD (PBCD) pipeline, radiation hits
in BCD images are detected and excluded, and BCD frames are then
combined into final PBCD mosaics. The detailed data reductions in the pipelines
can be found in the IRAC Data Handbook (version 3.0; \citealt{rsg+06}). 
For our 2007 data, the target's nearby region in the 4.5 $\mu$m 
images suffered the ``column pull-down" effect, an artifact in which a column 
of the detector array has a reduced intensity when a place in the column 
reaches a level of approximately 35000 digital numbers (DNs) due to a bright 
star or cosmic rays.  We corrected the BCD images for the artifact by using 
the tool {\tt cosmetic.pl}. The artifact-corrected BCD images were then 
combined into PBCD mosaics by using the tool {\tt mosaic.pl}. Both tools 
are provided by SSC in the {\tt MOPEX} package. In our PBCD data processing, 
the latest June 2006 permanently damaged pixels mask (pmask) image, 
provided by SSC, was used.

The target field in the 4.5 $\mu$m PBCD images is crowded. 
We used {\tt DOPHOT} \citep{sms93}, a point-spread function 
(PSF) fitting program, for photometry. 
For the 8.0 $\mu$m images, in
which the sky background dominates in brightness and varies substantially
over the source field region, we found that aperture photometry provided
more consistent measurements (this was also suggested by the \spz\ Helpdesk).
The program {\tt phot} in the IRAF package {\tt apphot} was used for
photometry. The aperture radius was 3 pixels and background 
annulus radii were 3--7 pixels. The {\tt DOPHOT} results
were also corrected to this aperture set \citep{rsg+06}. The correction factor
was derived by comparing the brightnesses resulting from 
the PSF-fitting photometry to those from the aperture photometry 
of a few well exposed, isolated stars in the source field. 

In addition, to eliminate systematic variations among the PBCD images 
that might be caused by changes such as in the instrument, observing 
conditions, or data reduction, we applied a differential photometry 
technique: we calibrated the brightness of our target with those of 
an ensemble of bright, non-variable stars in the images. The base images were 
the 2005 data, which have longer exposure times and thus higher 
signal-to-noise ratios.
\begin{center}
\includegraphics[scale=0.72]{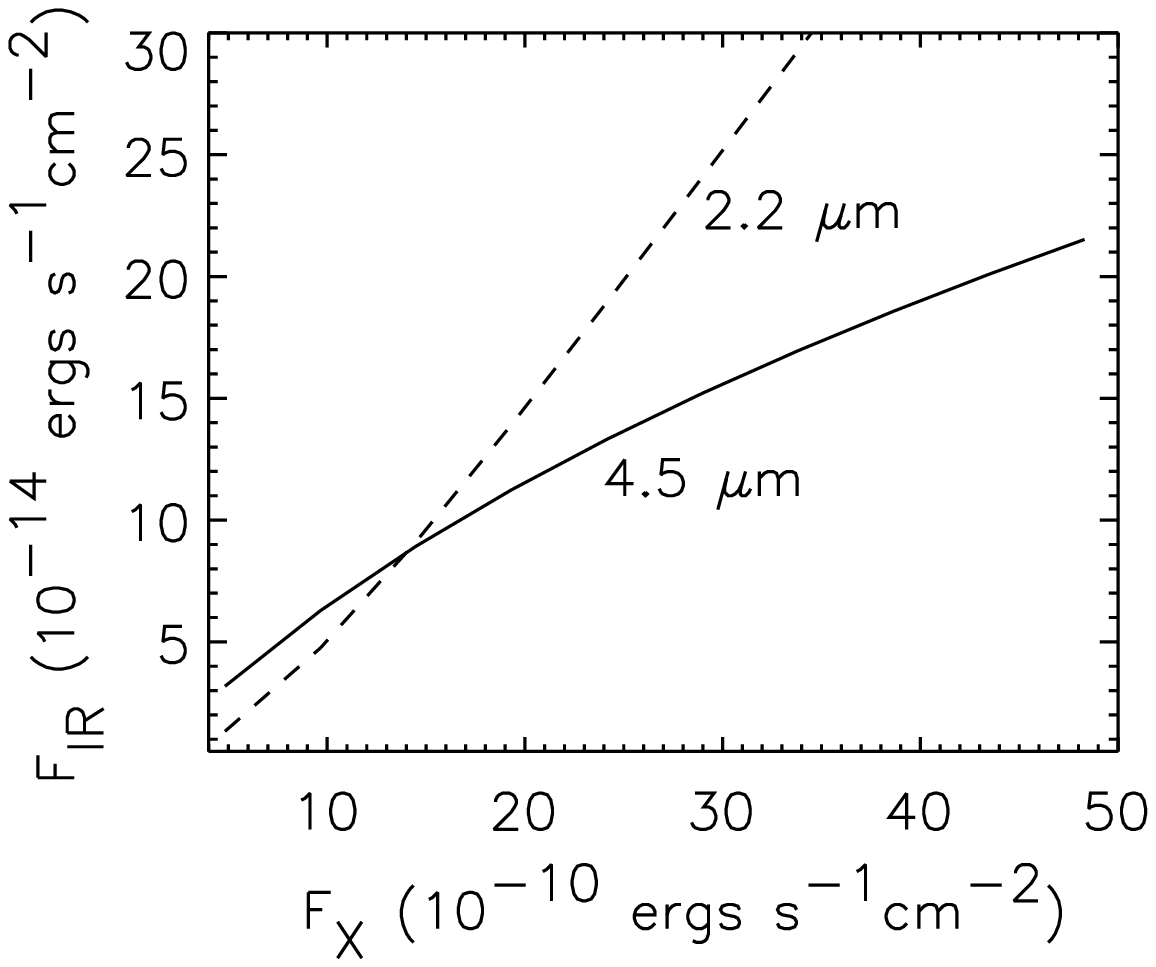}
\figcaption{The 4.5 $\mu$m flux 
(solid curve) as a function of the X-ray flux (0.5--10 keV),
derived from the dust disk model for \axp\ \citep{wck06}.
Similar NIR $K$ band (2.2 $\mu$m) flux changes (dashed curve) are also
expected if the disk structure is assumed to be fixed.
\label{fig:disk}}
\end{center}

\section{RESULTS}
\label{sec:res}

The flux density measurements are given in Table~\ref{tab:obs}. As can be seen,
the flux densities are consistent with being a constant within the 
uncertainties.
No significant variations were detected from our observations.
Compared to the 2005 flux densities, the upper limits (90\% confidence) on the 
flux density
changes are 13\% at 4.5 $\mu$m and 38\% at 8.0 $\mu$m. We note that
the flux densities in Table~\ref{tab:obs} for the 2005 observation 
are slightly lower than  
those previously reported, 36.3$\pm$3.6 and 51.9$\pm$5.2 $\mu$Jy
at 4.5 and 8.0 $\mu$m, respectively \citep{wck06}. For the 4.5 $\mu$m results, 
the difference is mainly caused by blending with nearby stars. A few photons
from nearby stars appear within 2--3 pixels radii from the center of our 
target. When aperture photometry was used (in \citealt{wck06}), the extra 
photons were included as the target counts. 
This can be verified by comparing the measurements from aperture photometry
to those from PSF-fitting. For example, when an aperture radius of 3 pixels
is used, the ratio between the fluxes from the first and latter photometry 
is 1.42$\pm$0.09 for our 
target, while the average ratio is 1.214$\pm$0.008 for a few  
non-blended test stars in the field. When the aperture radius is changed to
2 pixels, the flux ratios for our target and the test stars are equal
(1.07$\pm$0.07 for the target versus 1.070$\pm$0.008 for the test stars).
For the 8.0 $\mu$m results, the difference, which is well within 
the uncertainties, is probably due to the different IRAC data 
pipelines [version S11.0 in \citet{wck06} versus S15.3.0 in this work].

We average the four 2007 4.5 $\mu$m fluxes, weighted by the flux
uncertainties, and find a mean of 32.1$\pm$1.2 $\mu$Jy. This
value is exactly the same as the 2005 result (note that the total exposure
time in each wavelength band of the 2007 observations is approximately 
the same as that of the 2005 observation). Adding a 5\% absolute 
calibration uncertainty in quadrature (\citealt{reach+05}; \citealt{rsg+06}), 
we obtain 32.1$\pm$2.0 $\mu$Jy. 
The 8.0 $\mu$m flux densities, as shown in Table~\ref{tab:obs}, 
have similar uncertainties, even
though the exposure time of the 2005 observation is four times that
of one 2007 observation. This is because the flux density uncertainties are 
dominated by the sky noise (e.g., \citealt{reach+05}) and there were 
intrinsic intensity variations within the sky annulus in the aperture 
photometry. Therefore,
the longer 2005 exposure time did not effectively reduce the sky noise.
We average the four flux densities and find an
average of 59.8$\pm$8.5 $\mu$Jy, where we use  
the uncertainty 8.0 $\mu$Jy (Table~\ref{tab:obs}) 
and add the 5\% absolute calibration uncertainty to it in quadrature.

\section{DISCUSSION}
\label{sec:disc}

There are a total of 11 reported $K$-band observations of \axp\ in the past 
9 years, and three of them were with simultaneous $JH$ observations
\citep{dv06,gon+07}. In the observations, the AXP exhibited
large and rapid flux variations in the $K$ band, and 
the $JH$ fluxes correlated with the $K$ but varied less.
Assuming that the $K$-band variations are sufficiently sampled,
the probability of seeing a $>$15\% flux change would be 60\%.
In one extreme case, a 40\% $K$ flux decrease occured within one day.
Our \spz\  observations were planned so that such strong variability 
would have been detected. Correlated flux changes between the MIR 
and $K$ bands are expected in the proposed dust disk model 
for \axp\ (Figure~\ref{fig:disk}; \citealt{wck06}), if 
X-ray flux changes are considered as the primary cause of the variability.

We did not detect the $K$-like flux variations in our MIR observations.
The \spz/IRAC fluxes were consistent with being a constant within
the uncertainties.  Given that the upper limit on the MIR 4.5 $\mu$m flux 
variations was $\lesssim 13$\%,  
the MIR emission variability was similar to the NIR $JH$ bands and less variable
than in the $K$ band. 
We note that in a $K$-band observation made one 
day prior to our first \spz\ observation, the source had 
$K=19.9\pm0.1$ mag \citep{gon+07}, approximately the average $K$ 
magnitude of \axp. Because the time scale of the $K$ flux variations
is not determined, this could suggest that 
the source was in a stable state in the NIR during our observations,
which could be a reason for the absence of MIR variability.
To further constrain the relation between the MIR and NIR emission,
frequent, simultaneous observations are needed.

If the MIR emission is truly less variable, 
the restricted $K$-band variability would demand explanation.  
Interestingly, similar NIR variability is seen in a few dust 
disk systems around young stars (e.g., \citealt{eir+02}). 
In those systems, the
variability has been suggested to be caused by structural changes 
of the inner disk. If this is the case for \axp, the properties in
the source's IR variability might be explainable. Since emission in 
the $K$ band 
would mainly arise from the inner edge of the disk and thus be
more sensitive to inner disk changes, it would vary more strongly than
emission in the $JH$ and \spz\ MIR bands.
It is not clear what would be the cause of the structural changes 
in the disk in \axp.  One possibility is variable X-ray heating, 
such as due to burst-related X-ray brightening. However, 
no significant short-term flux variations have been found 
in the X-ray observations of the source \citep{dv06},
and thus far only a few bursts have been found in bi-weekly
\textit{RXTE} X-ray monitoring observations \citep{dkgw06,gav+07},
indicating infrequency of the burst events. 
Thus in the absence of X-ray variability, the $K$-band variability
challenges the dust disk model \citep{dv06}. 

In addition, we did not detect any significant MIR flux changes from 
\axp\  in our \spz\ IRAC observations following the large X-ray burst. 
The 4.5/8.0 $\mu$m fluxes were nearly the same as those obtained in 2005.
As demonstrated by X-ray observations of the source that were made several 
days prior to our first \spz\ observation (Gonzalez et al. 2007), 
the AXP unfortunately had already returned to quiescence during our 
observations. 
Also as shown in the long-term X-ray monitoring observations, the quiescent 
X-ray flux of \axp\  has been nearly constant and stable over
the past decade (\citealt*{dkg07}; Gonzalez et al. 2007).  
Therefore, our non-detection is consistent with the source's state 
at X-ray energies at the time. 

\acknowledgements
We thank the {\em Spitzer} Science Center for granting us the observations
and the Helpdesk at {\em Spitzer} Science Center for help with data reduction
and analysis.
This research was supported by NSERC via a Discovery Grant
and by the FQRNT and cifar.  VMK holds a Canada Research Chair and
the Lorne Trottier Chair in Astrophysics \& Cosmology, and is a
R. Howard Webster Foundation Fellow of cifar.

\bibliographystyle{apj}

\clearpage
\begin{deluxetable}{ccc}
\tablewidth{0pt}
\tablecaption{\spz\ IRAC 4.5/8.0 $\mu$m flux density measurements of AXP \axp\label{tab:obs}}
\tablehead{
\colhead{Observation Start Time} & \colhead{$F_{4.5}$} & \colhead{$F_{8.0}$}\\
\colhead{(UTC)}			 & \colhead{($\mu$Jy)} & \colhead{($\mu$Jy)}}
\startdata
2005-01-17 22:31		 & 32.1$\pm$1.2	       & 48.8$\pm$7.6 \\ 
2007-02-14 05:50		 & 31.2$\pm$2.2	       & 58.2$\pm$8.8 \\
2007-02-15 17:59		 & 33.0$\pm$2.5        & 53.5$\pm$8.8 \\
2007-02-18 09:02		 & 32.8$\pm$2.2	       & 65.6$\pm$8.0 \\
2007-02-21 15:58		 & 31.5$\pm$2.3	       & 61.2$\pm$9.8 \\
\enddata
\tablecomments{5\% absolute calibration uncertainty is not included.}
\end{deluxetable}

\end{document}